\documentclass[twocolumn,showpacs,aps,prl,superscriptaddress]{revtex4b5}

\usepackage{graphicx}
\usepackage{dcolumn}
\usepackage{amsmath}
\usepackage{epsfig}

\input pubboard/babarsym

\newcommand{\BABARPubYear}    {01}
\newcommand{\BABARPubNumber}  {18}

\newcommand{\SLACPubNumber} {8904}

\def\figurebox#1#2#3{%
    \def\arg{#3}%
    \ifx\arg\empty
    {\hfill\vbox{\hsize#2\hrule\hbox to #2{\vrule\hfill\vbox to #1{\hsize#2\vfill}\vrule}\hrule}\hfill}%
    \else
    {\hfill\epsfbox{#3}\hfill}%
    \fi}

\long\def\inst#1{\par\nobreak\kern 4pt\nobreak
    {\it #1}\par\vskip 10pt plus 3pt minus 3pt}

\begin{document}


\begin{flushleft}
\babar-PUB-\BABARPubYear/\BABARPubNumber\\
SLAC-PUB-\SLACPubNumber\\
\end{flushleft}

\title{
\vskip 10mm
{\large \bf Observation of {\boldmath \CP} violation in the \Bz meson system } \begin{center} 
\vskip 10mm
{The \babar\ Collaboration}
\end{center}
}

%
\author{B.~Aubert}
\author{D.~Boutigny}
\author{J.-M.~Gaillard}
\author{A.~Hicheur}
\author{Y.~Karyotakis}
\author{J.~P.~Lees}
\author{P.~Robbe}
\author{V.~Tisserand}
\affiliation{Laboratoire de Physique des Particules, F-74941 Annecy-le-Vieux, France }
\author{A.~Palano}
\affiliation{Universit\`a di Bari, Dipartimento di Fisica and INFN, I-70126 Bari, Italy }
\author{G.~P.~Chen}
\author{J.~C.~Chen}
\author{N.~D.~Qi}
\author{G.~Rong}
\author{P.~Wang}
\author{Y.~S.~Zhu}
\affiliation{Institute of High Energy Physics, Beijing 100039, China }
% ...
\collaboration{The \babar\ Collaboration}
\noaffiliation

\date{July 5, 2001}

\begin{abstract}
We present an updated measurement of time-dependent \CP-violating asymmetries in neutral
$B$ decays with the \babar\ detector at the \pep2\ asymmetric \BF\ at SLAC.  
This result uses an additional sample of $\FourS$ decays collected in 2001, bringing
the data available to 32 million $B\Bbar$ pairs.
We select events in which one neutral $B$ meson is
fully reconstructed in a 
final state containing charmonium and
the flavor of the other neutral $B$ meson is determined from its decay products.
The amplitude of the \CP-violating asymmetry, which in the Standard Model is 
proportional to \stwob, is derived from the decay time distributions in such events.
The result $\stwob=0.59 \pm 0.14\ {\rm (stat)} \pm 0.05\ {\rm (syst)}$ establishes 
\CP violation in the \Bz\ meson system.
We also determine $\vert\lambda\vert = 0.93 \pm 0.09\ (\rm {stat}) \pm 0.03\ (\rm {syst})$, 
consistent with no direct \CP  violation.
\end{abstract}


\maketitle

\CP\ violation has been a central concern of particle physics since its
discovery in 1964 in the decays of \KL\ mesons\cite{KLCP}. To date, this 
phenomenon has not been observed in any other system.
An elegant
explanation of this  effect  was proposed by Kobayashi and Maskawa,
as a  \CP-violating phase in the three-generation 
CKM quark-mixing matrix~\cite{CKM}. 
In this picture, measurements of \CP-violating
asymmetries in the time distributions of  
\Bz decays to charmonium final states provide a direct test of the
Standard Model of electroweak interactions~\cite{BCP}, 
free of corrections from strong
interactions that obscure the theoretical interpretation of the observed
\CP\ violation in \KL\ decays.

Measurements of the \CP-violating asymmetry  parameter \stwob\ have recently
been reported by the \babar~\cite{BABARPRL} and Belle~\cite{BELLEPRL}
collaborations, from data taken in 1999-2000  at  the  PEP-II and 
KEKB  asymmetric-energy
\epem colliders respectively, with better precision than previous 
experiments~\cite{OPALCDFALEPH}.
In this Letter we report a new  measurement of
\stwob, enhanced by 9 million $B\Bbar$ pairs collected in 2001, 
additional decay modes, and  improvements in data reconstruction and analysis.
The \babar\ detector and  the experimental method are described in
Refs.~\cite{BABARNIM,BABARPRL}, so the discussion here is limited to items and issues pertinent
to the current analysis.

The complete data set (32 million  $B\Bbar$ pairs) has been used to 
fully reconstruct a sample $B_{\CP}$ of neutral $B$ mesons decaying to 
the $\jpsi\KS$, $\psitwos\KS$,  
$\jpsi\KL$, $\chicone\KS$, and 
$\jpsi\Kstarz (\Kstarz\to\KS\piz ) $ final states. 
The last two modes have been added since Ref.~\cite{BABARPRL}. 
There are several other significant changes in the analysis. 
Improvements in track and \KS\ 
reconstruction efficiency in 2001 data
produce an approximately 30\% increase in the yields for a given luminosity.
Better alignment of the tracking systems in 2001 data and improvements in the tagging vertex
reconstruction algorithm  increase the sensitivity of the measurement
by an additional 10\%. Optimization of the  $\jpsi\KL$ selection 
increases the purity of this sample.
The final $B_{\CP}$ sample contains about 640 signal events and, with all
the improvements, the statistical power of the analysis is almost doubled
with respect to  that of Ref.~\cite{BABARPRL}.

We examine each of the events in the  $B_{\CP}$ sample for
evidence that the other neutral $B$ meson
decayed as a \Bz or a \Bzb (flavor tag).
The decay-time distributions for events with a \Bz or a \Bzb tag can be expressed
in terms of a complex parameter $\lambda$ that depends on both
\BzBzb mixing 
and on the amplitudes describing \Bzb and
\Bz decay to a common final state $f$~\cite{lambda}.
The distribution
${\rm f}_+({\rm f}_-)$ of the decay rate 
when the
tagging meson is a $\Bz (\Bzb)$ is given by
\begin{eqnarray}
{\rm f}_\pm(\, \deltat) = {\frac{{\rm e}^{{- \left| \deltat \right|}/\tau_{\Bz} }}{2\tau_{\Bz}
(1+|\lambda|^2) }}  \times  \left[ \ {\frac{1 + |\lambda|^2}{2}} \hbox to 2cm{}
\right. \nonumber \\ 
\left. 
\pm {\ \mathop{\cal I\mkern -2.0mu\mit m}}
\lambda  \sin{( \Delta m_{B^0}  \deltat )} 
\mp { \frac{1  - |\lambda|^2 } {2} }  
  \cos{( \Delta m_{B^0}  \deltat) }   \right],
\label{eq:timedist}
\end{eqnarray}
where $\deltat=t_{CP}-t_{\rm tag}$ is the time between the two \B decays, 
$\tau_{\Bz}$ is the \Bz lifetime and $\Delta m_{B^0}$ is the mass difference determined
from \BzBzb mixing~\cite{PDG2000}.
The first oscillatory term in Eq.~\ref{eq:timedist} is due to interference between 
direct decay and decay after mixing. 
A difference between the \Bz and \Bzb \deltat distributions or
a \deltat asymmetry for either flavor tag is evidence for \CP violation.

In the Standard Model $\lambda=\eta_f e^{-2i\beta}$ for
charmonium-containing $b\to\ccbar s$ decays, $\eta_f$ is the \CP eigenvalue of
the state $f$ and
$\beta = \arg \left[\, -V_{\rm cd}^{}V_{\rm cb}^* / V_{\rm td}^{}V_{\rm tb}^*\, \right]$
is an angle of the Unitarity Triangle of the three-generation CKM matrix~\cite{CKM}.
Thus, the time-dependent \CP-violating asymmetry is
\begin{eqnarray}
A_{\CP}(\deltat) &\equiv&  \frac{ {\rm f}_+(\deltat)  -  {\rm f}_-(\deltat) }
{ {\rm f}_+(\deltat) + {\rm f}_-(\deltat) } \nonumber \\%
&=& -\eta_f \stwob \sin{ (\Delta m_{B^0} \, \deltat )} , 
\label{eq:asymmetry}
\end{eqnarray}
where $\eta_f=-1$ for $\jpsi\KS$, $\psitwos\KS$ and $\chicone \KS$ and
$+1$ for $\jpsi\KL$. Due to the presence of even (L=0, 2) and odd (L=1)
orbital angular momenta in the $\jpsi\Kstarz (\Kstarz\to\KS\piz) $ system, 
there can be \CP-even and \CP-odd contributions to the decay rate. 
When the angular information in the decay is ignored, the measured \CP\ asymmetry 
in $\jpsi\Kstarz$ is reduced by a dilution factor $D_{\perp} = 1-2R_{\perp}$, where
$R_{\perp}$ is the fraction of the L=1 component. We have measured 
$R_{\perp} = (16 \pm 3.5)\% $ ~\cite{BABARTRANS} which, after 
acceptance corrections, leads to an effective 
$\eta_f = 0.65 \pm 0.07$ for the $\jpsi\Kstarz$ mode. 

The hadronic event selection,
lepton and charged kaon identification, 
and $\jpsi$ and $\psitwos$ reconstruction 
relevant to this analysis have been described in Ref.~\cite{BABARPRL}, as have  
the selection criteria for the channels
$\jpsi\KS\ (\KS \to \pipi, \ppz)$,
$\psitwos\KS\ (\KS \to \pipi)$, and $\jpsi\KL$.
In the $\jpsi\KL$ selection, 
the transverse missing momentum requirement has been reoptimized
for the $A_{\CP}$ study.

For the decay $\Bz\to\chicone\KS$, the mode $\chicone \to \jpsi \gamma$ is 
reconstructed with mass-constrained \jpsi\ candidates 
selected 
as in other charmonium channels~\cite{BABARPRL}.
Photons must  
have an energy greater than 150\mev and must 
not be associated with any reconstructed \piz .  
The resulting $\jpsi\gamma$ mass is required to be within 
35\mevcc\ of the $\chicone$ mass~\cite{PDG2000}.

For the decay $\Bz\to\jpsi \Kstarz$, 
the $\Kstarz \to \KS\piz $ candidate is formed by combining a
$\piz \to \gamma \gamma$ candidate
satisfying $106 \leq m_{\gamma\gamma} \leq 153$\mevcc\ 
with a \KS\ candidate. 
The cosine of the angle between the \KS\ momentum vector 
in the \Kstarz\ rest frame and the \Kstarz\ 
momentum defined in the \B\ rest frame is required to be less than 0.95. 
We require $ 796 \leq m_{\KS\piz} \leq 996$\mevcc.

\begin{figure}[!]
\begin{center}
\epsfxsize8.6cm
\figurebox{}{}{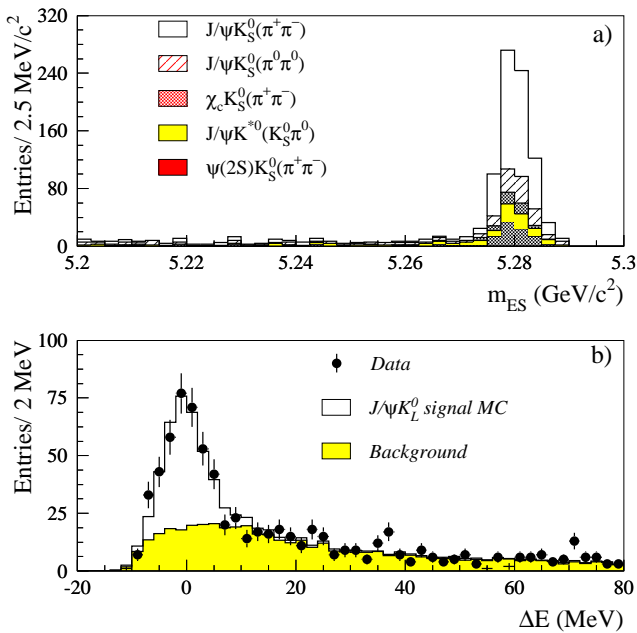}
\caption{ a) Distribution of \mes\ 
for $B_{\CP}$ candidates 
having a \KS 
in the final state;
b) distribution of $\Delta E$ for $\jpsi\KL$ candidates.}
\label{fig:prlfig1}

\end{center}
\end{figure}

$B_{\CP}$ candidates 
are selected by
requiring that the difference $\Delta E$ between their energy
and the beam energy in the center-of-mass frame be less than 
$3\sigma$ from zero. For modes involving \KS,
the beam-energy substituted mass 
$\mes=\sqrt{{(E^{\rm cm}_{\rm beam})^2}-(p_B^{\rm cm})^2}$ 
must be greater than $5.2\gevcc$. 
The resolution for $\Delta E$ is about 10\mev, except for
the $\KS\to\piz\piz$ mode (33\mev), the $\jpsi\Kstarz$ (20\mev ) and  
the $\jpsi\KL$ modes (3.5\mev after \B mass constraint).
For the purpose of determining numbers of events and purities, 
a signal region $\mes > 5.27\gevcc$ is used for
all modes except $\jpsi\KL$ and $\jpsi \Kstarz $.

Figure~\ref{fig:prlfig1} shows the resulting \mes distributions
for $B_{\CP}$ candidates containing a \KS and the $\Delta E$ distribution 
for the candidates containing a \KL.
The $B_{\CP}$ sample contains 1230 events in the signal region 
(before tag and vertex requirements), 
with an estimated background of 200 events, 
predominantly in the $\jpsi\KL$ channel. 
For that channel, the
composition, effective $\eta_f$, and $\Delta E$ distributions of the individual background sources
are taken either from a Monte Carlo simulation (for $B$ decays to \jpsi) or
from the $m_{\ell^+ \ell^-}$ sidebands in data.

A measurement of $A_{\CP}$ requires a determination of
the experimental \deltat resolution and  
the fraction of events in which the tag assignment is incorrect. 
A mistag fraction \mistag reduces the observed
asymmetry by a factor $(1-2\mistag)$.
A sample of self-tagging $B$ decays $B_{{\rm flav}}$
used in the determination of the mistag fractions 
and \deltat resolution functions
consists of the channels $D^{(*)-}h^+(h^+=\pi^+,\rho^+,a_1^+$) and 
$\jpsi K^{*0}\ (K^{*0}\to K^+\pi^-)$~\cite{conjugates}.
A control sample of
charged $B$ mesons decaying to the final states $\jpsi K^{(*)+}$, 
$\psitwos K^+$, $\chicone K^+$   
and $\Dbar^{(*)0}\pi^+$ is used for validation studies.

For flavor tagging, we exploit information from the 
other $B$ decay in the event. 
Each event is assigned to one of four hierarchical,
mutually exclusive tagging categories or excluded from further analysis. 
The {\tt Lepton} and {\tt Kaon} categories contain events with high momentum
leptons from semileptonic \B\ decays
or with kaons whose charge is correlated with the flavor of the decaying $b$ quark
({\it e.g.} a positive lepton or kaon yields a \Bz tag).
The {\tt NT1} and {\tt NT2} categories are based on a neural network algorithm whose
tagging power arises primarily from soft pions from \Dstarp decays and from recovering 
unidentified isolated primary leptons~\cite{BABARPRL}.

The numbers of tagged events and the signal purities, determined
from fits to the \mes (all \KS modes except \Kstarz )  or 
$\Delta E$ (\KL\ mode) distributions in data  or from 
Monte Carlo simulation (\Kstarz mode), are shown in Table~\ref{tab:result}. 
The efficiencies and mistag fractions for the four tagging categories are 
measured from data and summarized in Table~\ref{tab:TagMix:mistag}.

\begin{table}[!htb] 
\caption{ 
Number of tagged events, signal purity and result of fitting for \CP\ asymmetries in 
the full \CP sample and in 
various subsamples, as well as in the $B_{\rm flav}$ and charged $B$ control samples.  
Errors are statistical only.}
\label{tab:result} 
\begin{ruledtabular} 
\begin{tabular*}{\hsize}{ l@{\extracolsep{0ptplus1fil}} r c@{\extracolsep{0ptplus1fil}} D{,}{\ \pm\ }{-1} } 
 Sample  & $N_{\rm tag}$ & Purity (\%) & \multicolumn{1}{c}{$\ \ \ \stwob$}  
\\ \colrule

$\jpsi\KS$,$\psitwos\KS$,$\chicone\KS$   & $480$        & $96$       &  0.56, 0.15   \\ 
$\jpsi \KL$ $(\eta_f=+1)$                & $273$        & $51$       &  0.70, 0.34   \\
$\jpsi\Kstarz ,\Kstarz \to \KS\piz       $& $50$         & $74$       &  0.82,1.00  \\ 
\hline
 Full \CP\ sample                        & $803$        & $80$       &  0.59,0.14   \\ 
\hline
\hline
\multicolumn{4}{l}{$\jpsi\KS$, $\psitwos\KS$, $\chicone\KS$  only  $(\eta_f=-1)$ }  \\
\hline
$\ \jpsi \KS$ ($\KS \to \pi^+ \pi^-$)    & $316$        & $98$       &  0.45, 0.18  \\ 
$\ \jpsi \KS$ ($\KS \to \pi^0 \pi^0$)    & $64$         & $94$       &  0.70,0.50  \\ 
$\ \psi(2S) \KS$ ($\KS \to \pi^+ \pi^-$) & $67$         & $98$       &  0.47,0.42   \\
$\ \chicone \KS $                        & $33$         & $97$       &  2.59,$$^{0.55}_{0.67}$$ \\
\hline 
$\ $ {\tt Lepton} tags                   & $74$         &  $100$     &  0.54,0.29   \\ 
$\ $ {\tt Kaon} tags                     & $271$        &  $98$      &  0.59, 0.20    \\ 
$\ $ {\tt NT1} tags                      & $46$         &  $97$      &  0.67, 0.45    \\ 
$\ $ {\tt NT2} tags                      & $89$         &  $95$      &  0.10, 0.74   \\ 
\hline 
$\ $ \Bz\ tags                           & $234$        &  $98$      &  0.50, 0.22     \\ 
$\ $ \Bzb\ tags                          & $246$        &  $97$      &  0.61,0.22     \\ 
\hline\hline
$B_{\rm flav}$ non-\CP sample            & $7591$       & $86$       &  0.02,0.04     \\
\hline 
Charged $B$ non-\CP sample        & $6814$       & $86$       &  0.03,0.04     \\

\end{tabular*} 
\end{ruledtabular} 
\end{table}

\begin{table}[!htb] 
\caption
{ Average mistag fractions $\mistag_i$ and mistag differences $\Delta\mistag_i=\mistag_i(\Bz)-\mistag_i(\Bzb)$
extracted for each tagging category $i$ from the maximum-likelihood fit to the time distribution for the 
fully-reconstructed \Bz\ sample ($B_{\rm flav}$+$B_{\CP}$). The figure of merit for tagging is 
the effective tagging efficiency $Q_i = \eps_i (1-2\mistag_i)^2$, where $\eps_i$ 
is the fraction of events with a reconstructed tag vertex that are
assigned to the $i^{th}$ category. 
Uncertainties are statistical only. The 
statistical error on \stwob is proportional to $1/\sqrt{Q}$, where $Q=\sum Q_i$. } 
\label{tab:TagMix:mistag} 
\begin{ruledtabular} 
\begin{tabular*}{\hsize}{ l@{\extracolsep{0ptplus1fil}} D{,}{\ \pm\ }{-1} @{\extracolsep{10ptplus1fil}} c@{\extracolsep{0ptplus1fil}} D{,}{\ \pm\ }{-1} @{\extracolsep{0ptplus1fil}} D{,}{\ \pm\ }{-1} } 
Category     & \multicolumn{1}{c}{$\ \ \ \varepsilon$ (\%)} & $\mistag$ (\%) & \multicolumn{1}{c}{$\ \ \ \ \Delta\mistag$ (\%)} & \multicolumn{1}{c}{$\ \ \ Q$ (\%)}       \\ \colrule 
{\tt Lepton} & 10.9,0.3 & $8.9\pm 1.3$ & 0.9,2.2  &   7.4,0.5  \\ 
{\tt Kaon}   & 35.8,0.5 & $17.6\pm 1.0$ & -1.9,1.5 &  15.0,0.9  \\ 
{\tt NT1}    &  7.8,0.3 & $22.0\pm 2.1$ & 5.6,3.2  &   2.5,0.4  \\ 
{\tt NT2}    & 13.8,0.3 & $35.1\pm1.9$ & -5.9,2.7 &   1.2,0.3  \\  \colrule 
All          & 68.4,0.7 &              &          &  26.1,1.2  \\ 
\end{tabular*} 
\end{ruledtabular} 
\end{table} 
The time interval \deltat between the two $B$ decays is then determined
from the $\deltaz = z_{\CP}-z_{\rm tag}$ measurement, including 
an event-by-event correction for the direction
of the $B$ with respect to the $z$ direction in the $\FourS$ frame. 
$z_{\CP}$ is determined from the charged tracks that constitute the $B_{\CP}$ candidate.
The tagging vertex is determined by fitting the
tracks not belonging to
the $B_{\CP}$ (or $B_{\rm flav}$) candidate to a common vertex. 
The method employed is identical to our previous analysis except
for the addition of a constraint from knowledge of the beam spot location
and beam direction, which increases its efficiency to 97\%. 
This is incorporated through the addition of a pseudotrack 
to the tagging vertex, 
computed from the  $B_{\CP}$ ($B_{\rm flav}$) 
vertex and three-momentum, the
beam spot (with a vertical size of 10\mum )
and the \FourS momentum.
For 99\% of the reconstructed vertices the r.m.s. \deltaz resolution is 180\mum.
An accepted candidate must have a converged fit for the $B_{\CP}$ and $B_{\rm tag}$ vertices,
an error of less than 400\mum on \deltaz, and a measured $\vert \deltat \vert < 20 \ps$.
After tag and vertexing requirements about 640 signal events remain.

The \stwob measurement is made with a simultaneous
unbinned maximum likelihood fit to the \deltat 
distributions of the $B_{\CP}$ and $B_{\rm flav}$ tagged samples.
The \deltat\ distribution of the former is given by Eq.~\ref{eq:timedist}, with $|\lambda|=1$. 
The  $B_{\rm flav}$ sample evolves according to the known rate
for flavor oscillations in neutral $B$ mesons~\cite{PDG2000}. The amplitudes for $B_{\CP}$ asymmetries
and for $B_{\rm flav}$ flavor oscillations are reduced
by the same factor $(1-2\mistag)$ due to wrong tags. Both distributions are 
convolved with a common
\deltat resolution function and corrected for backgrounds.
Events are assigned signal and background probabilities based on the
\mes\ (all modes except $\jpsi\KL$) or $\Delta E$ ($\jpsi\KL$) distributions. 

The representation of the \deltat\ resolution function is the same as in
\cite{BABARPRL} with small changes: all offsets are modeled 
to be proportional to  $\sigma_{\Delta t}$, which is correlated with the weight 
that the daughters of long-lived charm
particles have in the tag vertex reconstruction. Separate resolution 
functions have been used for
the data collected in 1999-2000 and 2001, due to
the significant improvement in the silicon vertex tracker (SVT) alignment. 
The scale factor for the tail component
is fixed to the Monte Carlo value since it 
is strongly correlated with the other resolution function parameters.

A  total of 45 parameters are varied in the likelihood fit, including
\stwob (1), the average mistag fraction $\mistag$ and the 
difference $\Delta\mistag$ between \Bz\ and \Bzb\ mistags for 
each tagging category (8), parameters for the signal \deltat
resolution (16), and parameters for background time dependence (9), \deltat resolution (3) and mistag fractions (8).
The determination of the mistag fractions and signal \deltat resolution function is dominated
by the large $B_{\rm flav}$ sample. Background parameters are governed
by events with $\mes < 5.27\gevcc$.
As a result, the largest correlation between \stwob\ and any linear combination of the other free
parameters is only 0.13.
We fix $\tau_{\Bz}=1.548\ps$ and $\Delta m_{B^0}=0.472\,\hbar\ps^{-1}$~\cite{PDG2000}.

\begin{figure}[t!]
\begin{center}
\epsfxsize8.6cm
\figurebox{}{}{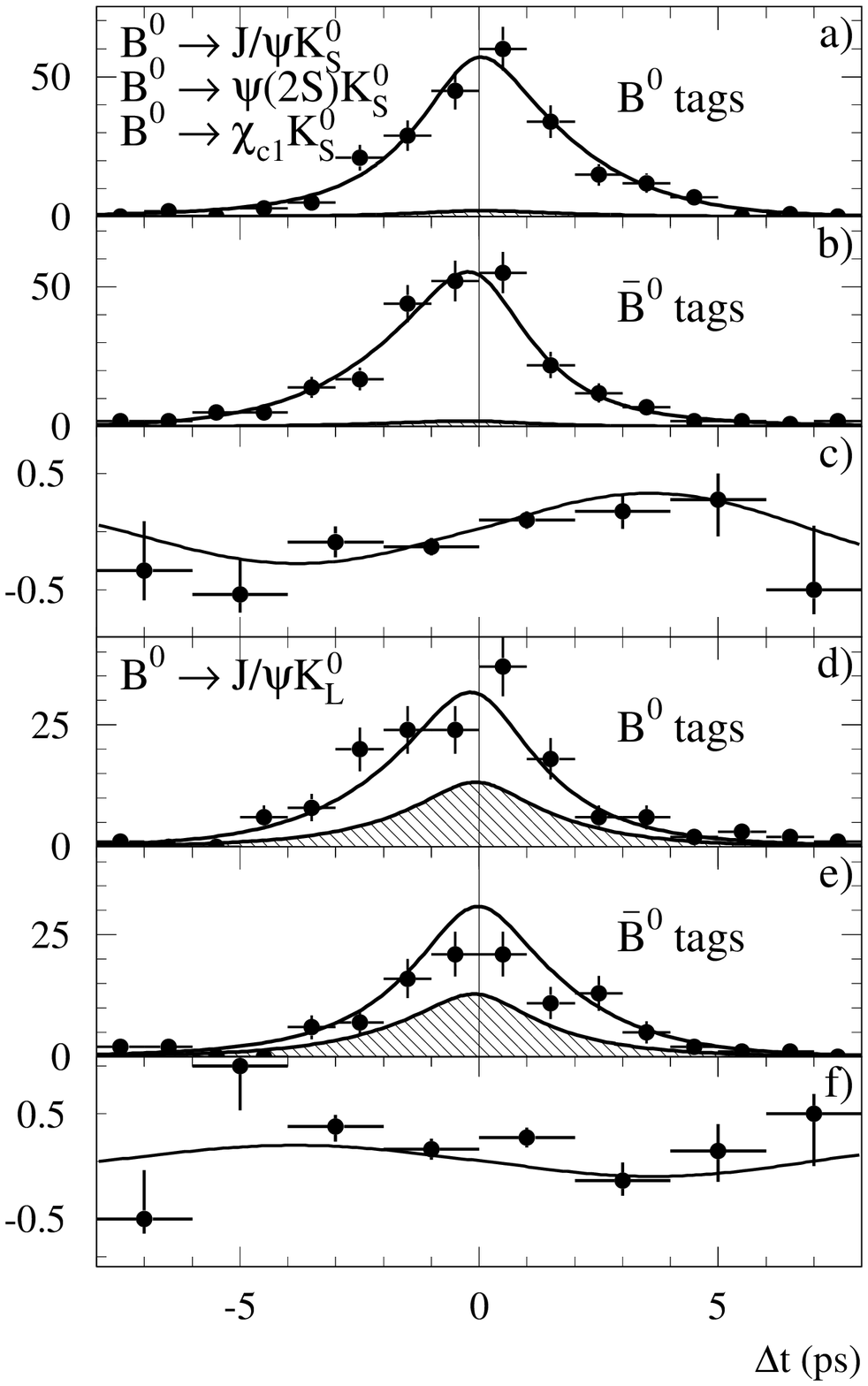}
\caption{Number of $\eta_f=-1$ candidates ($\jpsi\KS$, $\psitwos\KS$, and 
$\chicone\KS $) in the signal region  a) with a \Bz tag $N_{\Bz}$ 
and b) with a \Bzb tag $N_{\Bzb}$, and c) the asymmetry
$(N_{\Bz}-N_{\Bzb})/(N_{\Bz}+N_{\Bzb})$, as functions of \deltat . The
solid curves represent the result of the combined fit to all selected \CP events;
the shaded regions represent the background contributions.
Figures d)--f) contain the corresponding information for the 
$\eta_f=+1$ mode $(\jpsi\KL)$.
The likelihood is normalized to the total number of 
\Bz and \Bzb tags. The value of \stwob is independent of the individual normalizations
and therefore of the difference between the number of \Bz and \Bzb tags.
}
\label{fig:asymlike}
\end{center}
\end{figure}

Figure~\ref{fig:asymlike} shows the $\deltat$ distributions and 
${A}_{\CP}$ as a function of \deltat overlaid     
with the likelihood fit result for the $\eta_f = -1$ and $\eta_f = +1$ samples.
The probability of obtaining a lower likelihood, evaluated using a Monte
Carlo technique, is 27\%.
The simultaneous fit to all \CP decay modes and flavor decay modes yields
\begin{eqnarray}
\stwob=0.59 \pm 0.14\ \stat \pm 0.05\ \syst. \nonumber
\end{eqnarray}
Repeating the fit with all parameters fixed to their determined values except \stwob, we find
a total contribution of $\pm 0.02$ to the error on \stwob\ due to the combined statistical
uncertainties in mistag fractions, \deltat\ resolution and background parameters.
The dominant sources of systematic error are the parameterization of 
the \deltat\ resolution function (0.03),
due in part to residual uncertainties in 
SVT alignment, possible differences in the mistag fractions
between the $B_{\CP}$ and $B_{\rm flav}$ samples (0.03),  
and uncertainties in the level, composition, and \CP\ asymmetry of the background
in the selected \CP events (0.02).
The systematic errors from
uncertainties in $\Delta m_{\Bz}$ and $\tau_{\Bz}$ and from the
parameterization of the background
in the $B_{\rm flav}$ sample are small;
an increase of $0.02\,\hbar\ps^{-1}$ in the value for $\Delta m_{\Bz}$ 
decreases \stwob\ by 0.015.  

The large sample of reconstructed events allows a number of consistency
checks, including separation of the data by decay mode, tagging
category and $B_{\rm tag}$ flavor. The results of fits to these subsamples
are shown in Table~\ref{tab:result}. 
The consistency between various modes is satisfactory, the 
probability of finding a worse agreement being 8\%. 
The observed asymmetry in the number of \Bz (160) 
and \Bzb (113) tags in the $\jpsi\KL$ sample 
has no impact on the \stwob measurement.
Table~\ref{tab:result} also
shows results of fits to the samples of non-\CP decay modes, where no statistically
significant asymmetry is found. Performing the current analysis on the
previously published data sample and decay modes 
yields a value of
\stwob=0.32$\pm$0.18,  consistent with the published
value~\cite{BABARPRL}. For only these decay modes, the year 2001 data
yield \stwob=0.83$\pm$0.23, consistent with the 1999-2000 results at the 1.8$\sigma$ level.

If $\vert\lambda\vert$ is allowed to float in the fit to the
$\eta_f=-1$ sample, which has high purity and requires minimal assumptions on the 
effect of backgrounds, the value obtained is 
$\vert\lambda\vert = 0.93 \pm 0.09\ (\rm{\stat}) \pm 0.03\ (\rm{\syst})$. 
The sources of the systematic error in this measurement are the same as in the 
\stwob analysis.
The coefficient of the 
$\sin{( \Delta m_{B^0}  \deltat )}$ term in Eq.~\ref{eq:timedist}
is measured to be $0.56\pm 0.15$ (\rm{stat}).

The measurement of $\stwob=0.59 \pm 0.14\ \stat \pm 0.05\ \syst $ reported here 
establishes \CP violation in the \Bz meson
system at the $4.1\sigma$ level. 
This significance is computed from the sum in quadrature of 
the statistical and additive systematic errors.
The probability of obtaining this value or higher in the absence of \CP violation
is less than 
$3 \times 10^{-5}$. The corresponding probability for the 
$\eta_f=-1$ modes alone is $2 \times 10^{-4}$.
This direct measurement is in agreement with the range implied
by measurements and theoretical estimates of the 
magnitudes of CKM matrix elements~\cite{CKMconstraints}.  

We are grateful for the 
extraordinary contributions of our \pep2\ colleagues in
achieving the excellent luminosity and machine conditions
that have made this work possible.
The collaborating institutions wish to thank 
SLAC for its support and the kind hospitality extended to them. 
This work is supported by
DOE
and NSF (USA),
NSERC (Canada),
IHEP (China),
CEA and
CNRS-IN2P3
(France),
BMBF
(Germany),
INFN (Italy),
NFR (Norway),
MST (Russia), and
PPARC (United Kingdom). 
Individuals have received support from the Swiss NSF, 
A.~P.~Sloan Foundation, 
Research Corporation,
and Alexander von Humboldt Foundation.

\end{document}